# Wideband Sensing and Optimization for Cognitive Radio Networks with Noise Variance Uncertainty

Tadilo Endeshaw Bogale, *Member, IEEE*, Luc Vandendorpe, *Fellow, IEEE* and Long Bao Le, *Senior Member, IEEE*

*Abstract*— This paper considers wide-band spectrum sensing and optimization for cognitive radio (CR) networks with noise variance uncertainty. It is assumed that the considered wide-band contains one or more white sub-bands. Under this assumption, we consider throughput maximization of the CR network while appropriately protecting the primary network. We address this problem as follows. First, we propose novel ratio based test statistics for detecting the edges of each sub-band. Second, we employ simple energy comparison approach to choose one reference white sub-band. Third, we propose novel generalized energy detector (GED) for examining each of the remaining sub-bands by exploiting the noise information of the reference white sub-band. Finally, we optimize the sensing time ($T_o$) to maximize the CR network throughput using the detection and false alarm probabilities of the GED. The proposed GED does not suffer from signal to noise ratio (SNR) wall and outperforms the existing signal detectors. Moreover, the relationship between the proposed GED and conventional energy detector (CED) is quantified analytically. We show that the optimal $T_o$ depends on the noise variance information. In particular, with 10TV bands, SNR=−20dB and 2s frame duration, we found that the optimal $T_o$ is 28.5ms (50.6ms) with perfect (imperfect) noise variance scenario.

## I. INTRODUCTION

Cognitive radio (CR) is one of the promising approaches to improve the spectral efficiency of current wireless networks [1], [2]. One key feature of a CR network is the potential to learn its surrounding radio environment, which is performed by the spectrum sensing (signal detection) part of the CR device. The most widely known spectrum sensing algorithms are matched filter, energy and cyclostationary based algorithms. Among these algorithms, matched filter is optimal, which, however, requires perfect synchronization between the primary transmitter and cognitive device [3]. The energy detector (hereafter it is referred as conventional energy detector (CED)) does not require any information about the primary user and it is simple to implement. However, the CED is very sensitive to the noise variance uncertainty, and there is a signal to noise ratio (SNR) wall below which this detector can not guarantee the desired detection performance [3]–[5]. Cyclostationary based detection algorithm is robust against noise variance uncertainty and it can reject the effect of adjacent channel interference, which unfortunately has high computational complexity and is sensitive to cyclic frequency mismatch [5]–[7].

In [8], the eigenvalue decomposition (EVD)-based signal detector is proposed. This detector is robust against noise variance uncertainty [9]. However, this detector is not able to achieve the desired performance in the presence of very small adjacent channel interference (ACI) signals [10]. Recently in [11], new Max-Min SNR based signal detector which employs linear combination approach of the oversampled received signal is proposed. Under noise variance uncertainty, simulation results demonstrate that this detector achieves better performance than those of the CED and EVD-based detectors in additive white Gaussian noise (AWGN) and Rayleigh fading channels. The detector of [11] also guarantee the desired probability of detection (false alarm) $P_d(P_f)$ in the presence of low (moderate) ACI signals. The main drawbacks of [11], however, are that the cognitive device requires accurate knowledge of the primary transmitter pulse shaping filter and rolloff factor. Furthermore, the approach of [11] employs oversampling of the received signal beyond the Nyquist rate which may not be desirable in practice as such operation requires expensive high-speed analog to digital converter. In addition, the theoretical $P_f$ and $P_d$ expressions are obtained by employing numerical methods. Nonetheless, as will be clear later, the detection algorithm of the current paper achieves better performance than that of the algorithm of [11] under similar assumptions. Furthermore, the proposed detector of the current paper requires neither the transmitter pulse shaping filter nor oversampling of the received signal. And the $P_f$ and $P_d$ expressions are derived in closed forms.

In [12] and [13], edge detection followed by CED approach is proposed for sensing wide-band signals. However, for any given SNR, the approach of these papers does not provide analytical method to examine the performance of their edge detector. And most importantly, for each sub-band, the work of these papers apply the threshold values of the CED by assuming that the smallest average energy of the sub-bands is equal to the true noise variance. However, as will be detailed later, if the noise variance is estimated from a given finite white sub-band, one can not directly apply the threshold of the CED. In [14], the throughput optimization problem for CR network has been considered by employing CED. In [15], multi-band joint detection approach is applied to maximize the throughput of wide-band CR network. This paper considers $K \geq 2$ non-overlapping frequency sub-bands each with predefined bandwidth and the optimal detection threshold was computed jointly for all sub-bands. The problem of designing the optimal sensing time and power allocation strategy for maximizing the ergodic throughput of a wide-band CR network is studied in [16]. This paper examines its problem for sensing based spectrum sharing and opportunistic spectrum access schemes. And to protect the primary users from harmful interference, the CED approach is employed. For each sub-band, the works of these papers employ the CED spectrum sensing approach. However, as we have explained previously, the CED suffers from SNR wall. Thus, it may not be possible to utilize the CED to maximize the throughput of the CR network for very low SNR of the primary transmitted signal (e.g., $-20$dB for the wireless regional area network (WRAN) as in [8], [17]). Recently, we have examined the throughput maximization problem in [18] for CR networks under noise variance uncertainty. This work considers the scenario where there are two sub-bands with known bandwidths. Furthermore, the second sub-band contains noise only information and is assumed to be known a priori.

In the current paper, we consider more general and practical scenario where we may have two or more sub-bands, and the bandwidths of these sub-bands are not known a priori. For such settings, we consider spectrum sensing and throughput optimization for wide-band CR networks with noise variance uncertainty[1]. It is assumed that the considered wide-band contains one or more white sub-bands. This assumption is reasonable since according to the Federal Communications Commission (FCC) report, spectrum utilization on most available frequency bands is quite low [1]. Moreover, we assume that the CR network performs sensing and transmission repeatedly over equal frame intervals. This frame-based sensing and transmission strategy has been commonly adopted in the literature [14]. This frame interval can be set to the required channel evacuation time which is 2s in the 802.22 standard for example [17]. Under these assumptions, we come up with the following key contributions.

- We propose a unified spectrum sensing and throughput optimization framework for wide-band CR networks. The proposed framework comprises of edge detection to determine different sub-bands, reference sub-band isolation to determine the white sub-band (i.e., a sub-band which contains noise only signal), generalized energy detection for each of the remaining sub-bands (i.e., the target sub-bands) based on the noise information collected from the reference white sub-band, and throughput maximization of all sub-bands. The generalized energy detector (GED) is designed to ensure the prescribed detection and false alarm pair without experiencing any SNR wall.
- We derive the $P_d$ and $P_f$ expressions for the proposed edge and generalized energy detectors. These derivations reveal that the detection performance of the proposed GED depends on the bandwidths of the reference white sub-band and the target sub-band. Moreover, the relation-ship between CED and the proposed GED is quantified analytically.
- We formulate the throughput maximization problem for wide-band CR network as a concave maximization problem where its solution is obtained by using convex optimization tools. We show that the optimal sensing time that maximizes the throughput using the proposed GED ($T_{oGED}$) is different from that of the CED ($T_{oCED}$). In other words, the optimal sensing durations with perfect and imperfect noise variances are not the same.
- Extensive numerical studies are conducted to investigate the performance of the proposed sensing and optimization framework. Specifically, we validate the analytical results by computer simulation. We demonstrate that the proposed detection algorithms are robust against noise variance uncertainty, and the proposed GED outperforms existing signal detectors. In an exemplifying setting with 10TV bands, SNR=$-20$dB, and 2s frame duration, we found that the optimal sensing times that maximize the throughput of the CR network are $T_{oGED} = 28.5$ms and $T_{oCED} = 50.6$ms.

[1]A wireless communication system is termed as wide-band when the transmitted signal's bandwidth is higher than the coherence bandwidth of the channel. In a typical communication environment, a signal with bandwidth in the order of few MHzs can be considered as wide-band (for example, a digital video broadcasting (DVB) signal of bandwidth 5MHz). However, in the current paper, we are examining a considerably larger bandwidth to better exploit the spectral opportunities for CR network transmission.

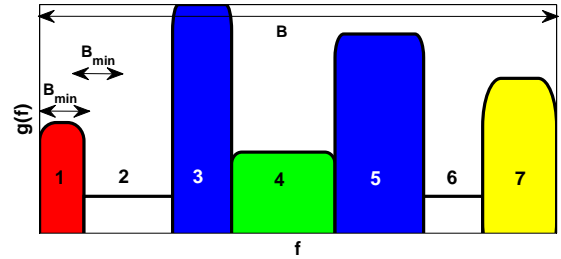

Fig. 1. The PSD of wide-band signal: $SB_2$ and $SB_6$ are white spaces.

The remaining part of this paper is organized as follows: Section II discusses the system model and the problem statement. The proposed ratio based edge detection, reference white sub-band detection and generalized energy detection algorithms are discussed in Sections III and IV. The sensing time optimization algorithm is presented in Section V. In Sections VI and VII, numerical and simulation results are presented for several practically relevant parameter settings. Finally, conclusions are drawn in Section VIII.

*Notations:* The following notations are used: $AE(.)$ denotes the average energy and $\lfloor x \rfloor$ ($\lceil x \rceil$) is the nearest integer less (greater) than or equal to $x$. The representations $SB$, s.t, $Pr(.)$, $(.)^\star$, $\mathrm{E}\{.\}$ and $|.|$ denote sub-band, subject to, probability, optimal, expectation and absolute value, respectively.

## II. SYSTEM MODEL AND PROBLEM STATEMENT

Consider a wide-band CR network that operates on the spectrum of $B$Hz where different sub-bands have different power spectral densities (PSDs). Fig. 1 illustrates a typical utilization pattern of the spectrum where the number of sub-bands is 7. We assume that a cognitive device attempts to identify and perform sensing for each sub-band and utilizes only white sub-bands for communications[2].

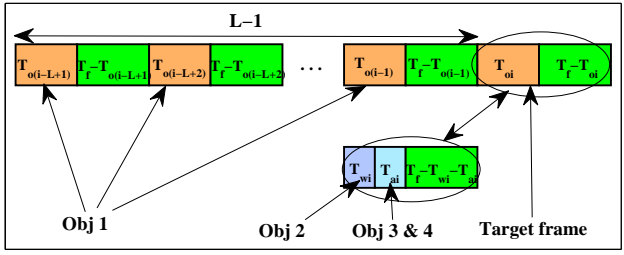

Fig. 2. The frame structure of a cognitive radio network.

The frame structure of a CR network is illustrated in Fig. 2 which shows $L$ consecutive frames. Here, each frame has a duration of $T_f = T_o + (T_f - T_o)$, where $T_o$ is used to sense each of the sub-bands and the remaining time $T_f - T_o$ is used for CR transmission. The sensing time $T_o$ is required to ensure that the primary network is sufficiently protected. This protection level is usually expressed in terms of the $P_d$ of each sub-band. The time $T_f - T_o$ is usually chosen to ensure that the white spaces are efficiently exploited. In this paper, we are interested in designing the spectrum sensing to maximize the throughput of the CR network while appropriately protecting the primary network. This problem can be expressed as

$$\max Th_f, \quad \text{s.t } P_{df}(\text{each sub band}) \geq \tilde{P}_d \quad (1)$$

where $Th_f$, $P_{df}(.)$ and $\tilde{P}_d$ are the total throughput of all target sub-bands, detection probability in each sub-band and target $P_d$ in frame $f$, respectively. To this end, we impose the following assumptions.

**Assumptions**:
1) The sampling frequency of the cognitive device is equal to $B$ (i.e., Nyquist sampling). The case of sub-Nyquist sampling approach has been discussed in Section VI.
2) We have considered multiple primary transmitters and one cognitive device (for sensing), and one secondary transmitter and receiver[3]. The channels between the primary transmitters and cognitive device, and the secondary transmitter and receiver are AWGN.
3) In each frame, the optimal sensing time is $T_o$ (i.e., the CR achieves maximum throughput at $T_o$).
4) The noise variance of each sub-band does not vary in each frame but can vary from one frame to another. In addition, the noise variance is not known perfectly.
5) The edges of each sub-band remain unchanged for $L_p$ consecutive frames[4]. However, in these frames, the PSD of each sub-band may change from one frame to another (i.e., a sub-band may contain noise only or signal plus noise in two consecutive frames).
6) In each frame, the examined band contains one or more white sub-bands and we will choose one of them as a reference white sub-band. However, the location of this sub-band is unknown a priori.
7) The bandwidth of each sub-band is at least $B_{min}$, and the number of sub-bands ($S_{sb}$) satisfies $2 \leq S_{sb} \leq S_{max}$, where $S_{max}$ is known but $S_{sb}$ is not known. Hence, each sub-band can have a bandwidth greater or equal to $B_{min} = \frac{B}{S_{max}}$.
8) The edge, reference white sub-band and generalized energy detectors are designed for a predefined minimum target SNRs ($\gamma_{min}$). In other words, these detectors can not guarantee the theoretical performance, if the SNR of the primary transmitted signal is less than $\gamma_{min}$. For example, $\gamma_{min}$ could be set to $\gamma_{min} = -20$dB.

Given these assumptions, (1) can be examined by addressing the following objectives for each frame.

**Obj 1**: Detecting the edges of each sub-band based on the received signal samples of the previous $L-1$ frames where $L < L_p$.

**Obj 2**: Determining the reference white sub-band (i.e., the sub-band which contains noise only). This step is required to get the noise information from the reference white sub-band.

**Obj 3**: Detecting each of the target sub-bands (i.e., sub-bands other than the reference white sub-band) using the proposed GED[5] by exploiting the noise information obtained from **Obj 2**.

**Obj 4**: Optimizing the sensing time to maximize the throughput of the CR network by employing the $P_d$ and $P_f$ expressions of the GED.

Fig. 2 shows how the frames and different parts in each frame are utilized to accomplish each of these objectives for any target frame (i.e., frame $i$ of the figure). As we can see from this figure, **Obj 1** is accomplished by utilizing the received signal samples of the previous $L-1$ frames, **Obj 2** is accomplished by using the received samples in the target frame of duration $T_w$ ($T_w$ depends on the output of **Obj 1**), and **Obj 3** and **Obj 4** are accomplished by collecting additional signal samples in the duration $T_a$ of the target frame ($T_a$ depends on the outputs of **Obj 1** and **Obj 2**). From this explanation, we can understand that **Obj 1** and **Obj 2** must be accomplished almost without any error to avoid the effect of error propagation. Detailed design to achieve these objectives is given in the following sections. As will be clear later, the proposed edge detection and reference white sub-band detection algorithms will address the requirements of **Obj 1** and **Obj 2**.

## III. EDGE AND REFERENCE WHITE SUB-BAND DETECTIONS

In this section, we discuss the proposed algorithms for edge and reference white sub-band detections.

---

[2]A CR network is a network that does not have exclusive right to use this wide-band. It is always termed as a secondary network.

[3]Note that the detected white sub-bands can be shared to more than one secondary user. However, this multiuser setup requires resource allocation approach. And extending the algorithms of the current paper for multiuser secondary network setup with resource allocation is beyond the scope of the paper and is still an open research topic.

[4]This assumption is reasonable as boundaries of the sub-bands would not change very quickly. Therefore, $L_p$ can be quite large in practice.

[5]As will be detailed later, our GED is not a straightforward extension of the CED.

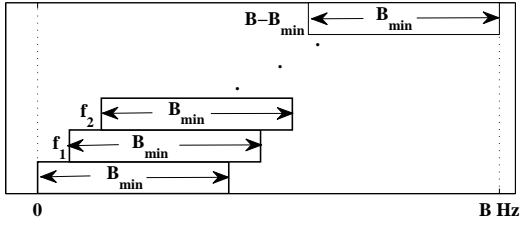

Fig. 3. Slided frequency regions for the edge detection algorithm.

## A. Ratio based Edge Detection

In general the edges of a sub-band can be characterized by the PSD difference between itself and its neighboring sub-band. Apparently, this fact is also seen from Fig. 1. Therefore, as the bandwidth of each sub-band is at least $B_{min}$, the sub-band edges can be determined if one compares the average energy in two halves of the desired frequency region where each half has size $\frac{B_{min}}{2}$.

To illustrate the brief idea behind our proposed edge detector, let us consider two frequency regions around sub-band 1 where each of them has size $B_{min}$ as shown in Fig. 1 where one frequency region coincides with sub-band 1 while the other frequency region has the falling edge of sub-band 1 at its center. Specifically, let us compute the following ratios for these two frequency regions as follows:

$$\tilde{T}_e = \frac{\text{AE}([0:B_{minh}])}{\text{AE}([B_{minh}:B_{min}])} \approx 1,$$
$$\tilde{T}_e = \frac{\text{AE}([B_{minh}:B_{min}])}{\text{AE}([B_{min}:B_{min}+B_{minh}])} \neq 1 \quad (2)$$

where $B_{minh} = \frac{B_{min}}{2}$ and $\text{AE}([x:y])$ denotes the average energy of the frequency region $[x:y]$. This expression implies that detecting the edges of each sub-band is equivalent to checking whether $\tilde{T}_e = 1$ (no edge) or $\tilde{T}_e \neq 1$ (there is an edge).

From this explanation, we can understand that all the edges of the wide-band can be identified using the aforementioned idea by sliding the desired frequency region of size $B_{min}$ as illustrated in Fig. 3 where the offset between two consecutive examined frequency regions can be set according to the resolution of the discrete time Fourier transform (DFT) of received time-domain signal samples, which will be described in more details later. In other words, the existence of edges in frequency regions $[f_i : f_i + B_{min}], \forall f_i \in [0, B - B_{min}]$ can be verified by computing $\tilde{T}_e$ where the frequency resolution $f_i - f_{i-1}$ can be set according to the DFT resolution.

In the following, we present the proposed edge detection scheme and its performance analysis in detail. Without loss of generality, we examine the frequency region $[0 : B_{min}]$ since the same technique can be applied for other frequency regions (see Fig. 3). We assume that the cognitive device collects signal samples of duration at least $T_{te}$ seconds in each frame. The number of sub-bands ($S_{sb}$) and bandwidth of each sub-band $\{B_j\}_{j=1}^{S_{sb}}$ are determined by employing the received signal samples of the previous $L-1$ frames (see Fig. 2). The base-band received signal of each frame $r(t)$ is expressed as[6]

$$r(t) = s(t) + w(t), \quad 0 \leq t \leq T_{te} \quad (3)$$

where $s(t)$ and $w(t)$ are the signal and noise components, respectively. By sampling this signal with period $\frac{1}{B}$, the sampled version of $r(t)$, with slight abuse of notation, can be expressed as

$$r[n] = s[n] + w[n], \quad n = 1, \cdots, N_e \quad (4)$$

where $N_e = \lfloor T_{te}B \rfloor$ is the number of received samples in $T_{te}$. It is assumed that $w[n], \forall n$ are independent and identically distributed (i.i.d) zero mean circularly symmetric complex Gaussian (ZMCSCG) random variables all with variance $\sigma^2$ which is unknown or known imperfectly. The DFT of $r[n]$ is given as [19]

$$\tilde{r}[m] = \sum_{n=1}^{N_e} \frac{r[n]\exp^{\frac{-i2\pi(m-1)(n-1)}{N_e}}}{\sqrt{N_e}}, \quad m = 1, \cdots, N_e. \quad (5)$$

By defining $\Delta f \triangleq \frac{B}{N_e} = f_i - f_{i-1}$, the PSD $g(f)$ of Fig. 1 and $\tilde{r}[m]$ are related as

$$g(f) \triangleq |\tilde{r}[m]|^2, \quad (m-1)\Delta f \leq f \leq m\Delta f, \quad \forall m.$$

Therefore, the average energies of the frequency regions $[0 : B_{minh}]$ and $[B_{minh} : B_{min}]$ of $g(f)$ are

$$\text{AE}([0:B_{minh}]) = \sum_{m=1}^{N_{eh}} \frac{|\tilde{r}[m]|^2}{N_{eh}} \triangleq \sum_{j=1}^{N_{eh}} \frac{|v[j]|^2}{N_{eh}} \triangleq \hat{M}_v$$

$$\text{AE}([B_{minh}:B_{min}]) = \sum_{m=1}^{N_{eh}} \frac{|\tilde{r}[N_{eh}+m]|^2}{N_{eh}}$$
$$\triangleq \sum_{j=1}^{N_{eh}} \frac{|\tilde{v}[j]|^2}{N_{eh}} \triangleq \hat{M}_{\tilde{v}} \quad (6)$$

where $N_{eh} = \lfloor B_{minh}T_{te} \rfloor$, $\{v[j] = \tilde{r}[m]\}_{j=m=1}^{N_{eh}}$ and $\{\tilde{v}[j] = \tilde{r}[N_{eh}+m]\}_{j=m=1}^{N_{eh}}$.

If the frequency region $[0 : B_{min}]$ does not contain any edge, $v[j]$ and $\tilde{v}[j]$ can be characterized by the same statistical parameter whereas, if this band contains an edge, $v[j]$ and $\tilde{v}[j]$ can not be characterized by the same statistical parameter. In a practical CR, $N_{eh}$ is in the order of thousands; thus, one can apply the well-known central limit theorem to model $v[j]$ and $\tilde{v}[j]$. Let $H_{0e}$, $H_{1e1}$ and $H_{1e2}$ represent the hypotheses in which there is no-edge, one falling edge and one rising edge in the examined frequency region $[0 : B_{min}]$, respectively. Then,

$$v[j] \sim \mathcal{N}(0, \sigma^2), \quad \tilde{v}[j] \sim \mathcal{N}(0, \sigma^2), \quad \text{Under } H_{0e}$$
$$v[j] \sim \mathcal{N}(\sqrt{\gamma_e}, \sigma^2), \quad \tilde{v}[j] \sim \mathcal{N}(0, \sigma^2), \quad \text{Under } H_{1e1}$$
$$v[j] \sim \mathcal{N}(0, \sigma^2), \quad \tilde{v}[j] \sim \mathcal{N}(\sqrt{\gamma_e}, \sigma^2), \quad \text{Under } H_{1e2} \quad (7)$$

where $\gamma_e$ is the SNR of $v[j](\tilde{v}[j])$ under $H_{1e1}(H_{1e2})$. Recall that the corresponding edge is at the center of the examined frequency region under the $H_{1e1}(H_{1e2})$ hypotheses due to the

---

[6]As will be clear in the sequel, $T_{te} \geq \frac{3.89 \times 10^5}{B}$ and $L \leq 60$ for practically relevant parameter settings. For example, when $B$=60MHz (i.e., 10 TV bands), we show that $T_{te} = 6.5$ms and $L = 55$.

aforementioned sliding procedure. For convenience, define the following terms

$$\hat{R}_{v\tilde{v}} \triangleq \frac{\hat{M}_v}{\hat{M}_{\tilde{v}}},$$

$$\tilde{R}_{v\tilde{v}} \triangleq \frac{N_{eh} \to \infty \frac{1}{N_{eh}} \sum_{j=1}^{N_{eh}} |v[j]|^2}{N_{eh} \to \infty \frac{1}{N_{eh}} \sum_{j=1}^{N_{eh}} |\tilde{v}[j]|^2} = \frac{M_v}{M_{\tilde{v}}} \quad (8)$$

where $M_v = N_{eh} \to \infty \frac{1}{N_{eh}} \sum_{j=1}^{N_{eh}} |v[j]|^2$ and $M_{\tilde{v}} = N_{eh} \to \infty \frac{1}{N_{eh}} \sum_{j=1}^{N_{eh}} |\tilde{v}[j]|^2$. It follows

$$\tilde{R}_{v\tilde{v}} = \begin{cases} 1, & \text{Under } H_{0e} \\ 1+\gamma_e, & \text{Under } H_{1e1} \\ (1+\gamma_e)^{-1}, & \text{Under } H_{1e2}. \end{cases} \quad (9)$$

To get the test statistics of the edge detector, we consider the following important theorem:

*Theorem 1*: Given a real valued function $\hat{R}_{v\tilde{v}} = \frac{\hat{M}_v}{\hat{M}_{\tilde{v}}}$, the asymptotic distribution of $\sqrt{N_{eh}}(\hat{R}_{v\tilde{v}} - \tilde{R}_{v\tilde{v}})$ is given by

$$\sqrt{N_{eh}}(\hat{R}_{v\tilde{v}} - \tilde{R}_{v\tilde{v}}) \sim \mathcal{N}(0, \tilde{\sigma}_e^2) \quad (10)$$

where $\tilde{\sigma}_e^2 = \mathbf{f}^T \mathbf{\Phi} \mathbf{f}$, $\mathbf{\Phi}$ is the asymptotic covariance matrix of a multivariate random variable $\sqrt{N_{eh}}([\hat{M}_v, \hat{M}_{\tilde{v}}]^T - [M_v, M_{\tilde{v}}]^T) \sim \mathcal{N}(\mathbf{0}, \mathbf{\Phi})$ and

$$\mathbf{f} = \left[\frac{\partial \hat{R}_{v\tilde{v}}}{\partial \hat{M}_v}, \frac{\partial \hat{R}_{v\tilde{v}}}{\partial \hat{M}_{\tilde{v}}}\right]^T_{\hat{M}_v = M_v, \hat{M}_{\tilde{v}} = M_{\tilde{v}}} = \left[\frac{1}{M_{\tilde{v}}}, -\frac{M_v}{M_{\tilde{v}}^2}\right]^T. \quad (11)$$

*Proof:* See *Theorem* 3. 3. A on page 122 of [20]. ∎

By employing this theorem, it can be shown that (see Appendix A for the proof)

$$\tilde{\sigma}_e^2 = \begin{cases} 2, & \text{Under } H_{0e} \\ 2(1+\gamma_e)^2, & \text{Under } H_{1e1} \\ 2(1+\gamma_e)^{-2}, & \text{Under } H_{1e2}. \end{cases} \quad (12)$$

Under $H_{0e}$ hypothesis, we have $\tilde{R}_{v\tilde{v}} = 1$ and $\tilde{\sigma}_e^2 = 2$. Moreover, the edges of each sub-band will remain constant for $L-1$ frames. Due to these reasons, we propose the edge detector with the test statistics

$$R_{eL} = \sum_{i=1}^{L-1} \tilde{R}_{ei}^2 \quad (13)$$

where

$$\tilde{R}_e = \sqrt{\frac{N_{eh}}{2}}\left(\frac{\hat{M}_v}{\hat{M}_{\tilde{v}}} - 1\right) \quad (14)$$

and $\tilde{R}_{ei}$ is $\tilde{R}_e$ of (14) corresponding to the $i$th frame. In the following, we provide the probability of false alarm ($P_{feL}$) and detection ($P_{deL}$) of this test statistics.

Under $H_{0e}$ hypothesis, it is clearly seen that $\tilde{R}_e \sim \mathcal{N}(0,1)$. Thus, $R_{eL}$ is a chi-square random variable with $L-1$ degrees of freedom. Hence, the false alarm probability is

$$P_{feL} = \Pr(R_{eL} > \lambda_e | H_{0e}) = \frac{\gamma(\frac{\lambda_e}{2}, \frac{L-1}{2})}{\Gamma(\frac{L-1}{2})} \quad (15)$$

where $\Gamma(.)$ is gamma function, $\gamma(.)$ is the incomplete gamma function [21] and $\lambda_e$ is the threshold.

Under $H_{1e}$ hypothesis (i.e., when there is an edge), we can model $\tilde{R}_e$ as

$$\tilde{R}_e \sim \mathcal{N}(\mu_{e1}, \tilde{\sigma}_{H_{1e1}}^2), \quad \text{Under } H_{1e1} \quad (16)$$

$$\tilde{R}_e \sim \mathcal{N}(\mu_{e2}, \tilde{\sigma}_{H_{1e2}}^2), \quad \text{Under } H_{1e2} \quad (17)$$

where $\mu_{e1} = \sqrt{N_{eh}}\gamma_e = \sqrt{T_{te}B_{minh}}\gamma_e$, $\tilde{\sigma}_{H_{1e1}} = 1 + \gamma_e$, $\mu_{e2} = -\sqrt{N_{eh}}\frac{\gamma_e}{\gamma_e+1} = -\sqrt{T_{te}B_{minh}}\frac{\gamma_e}{\gamma_e+1}$ and $\tilde{\sigma}_{H_{1e2}} = (1+\gamma_e)^{-1}$. Furthermore, we can express $R_{eL}$ as $\tilde{\sigma}_{H_{1e1}}^2 \tilde{R}_{eL1}$ and $\tilde{\sigma}_{H_{1e2}}^2 \tilde{R}_{eL2}$ under $H_{1e1}$ and $H_{1e2}$ hypotheses, respectively, where $\tilde{R}_{eL1}$ and $\tilde{R}_{eL2}$ are non-central chi-square random variables each with $L-1$ degrees of freedom, and non-central parameters $\mu_{eL1} = \mu_{e1}^2(L-1)$ and $\mu_{eL2} = \mu_{e2}^2(L-1)$, respectively. Hence, the detection probabilities under $H_{1e1}$ and $H_{1e2}$ hypotheses are

$$P_{deL1} = \Pr(R_{eL} > \lambda_e | H_{1e1}) \quad (18)$$
$$= \Pr\left(\tilde{R}_{eL} > \frac{\lambda_e}{\tilde{\sigma}_{H_{1e1}}^2}\bigg|H_{1e1}\right) = Q_{\frac{L-1}{2}}\left(\sqrt{\mu_{eL1}}, \frac{\sqrt{\lambda_e}}{\tilde{\sigma}_{H_{1e1}}}\right)$$

$$P_{deL2} = \Pr(R_{eL} > \lambda_e | H_{1e2}) \quad (19)$$
$$= \Pr\left(\tilde{R}_{eL} > \frac{\lambda_e}{\tilde{\sigma}_{H_{1e2}}^2}\bigg|H_{1e2}\right) = Q_{\frac{L-1}{2}}\left(\sqrt{\mu_{eL2}}, \frac{\sqrt{\lambda_e}}{\tilde{\sigma}_{H_{1e2}}}\right)$$

where $Q_M(a,b)$ is the Marcum Q-function [21]

Note that the result of the edge detector (i.e., a rising and falling edges) can provide some information about the PSDs of different sub-bands. However, the whiteness of each sub-band can not be determined just by applying the edge detection algorithm. To justify this fact, we consider two consecutive edges, one is a falling edge (e.g., edge 3) and the other is a rising edge (e.g., edge 4) of Fig. 1. As can be seen from these edges, the frequency range extending from a falling edge to a rising edge may not be necessarily a white sub-band. This is evidently confirmed by looking sub-band 4 of Fig. 1.

*B. Reference White Sub-band Detection*

Suppose that we have identified the edges of each sub-band (i.e., **Obj 1**). The next objective will be to reliably detect the reference white sub-band (i.e., **Obj 2**). We propose to choose the sub-band that has the *least average energy* to be the *reference white sub-band*. Moreover, if $SB_i$ is chosen as a reference white sub-band and $SB_k, k \neq i$ are other sub-bands that may contain signal plus noise with $\sum_{j=1}^{S_{sb}} B_j = B$, **Obj 2** is accomplished when we ensure $\text{AE}(SB_k) \geq \text{AE}(SB_i), \forall k \neq i$ with some predefined quality $\tilde{P}_{dwref}$ (e.g., $\tilde{P}_{dwref} = 0.999$ to avoid error propagation). Furthermore, as we can see from Fig. 2, to achieve **Obj 2**, the cognitive device must receive signal samples for the duration of $T_w$. Therefore, the aim of this subsection is to compute $T_w$ ensuring

$$P_{dwref} = Pr(\text{AE}(SB_k) \geq \text{AE}(SB_i)) \geq \tilde{P}_{dwref}, \quad \forall k \neq i. \quad (20)$$

In $T_w$ time duration, the cognitive device receives $N_w = \lfloor T_w B \rfloor$ samples. By denoting these samples as $\{u[j]\}_{j=1}^{N_w}$, the DFT of $u[j], \forall j$ is given as

$$\tilde{u}[j] = \sum_{m=1}^{N_w} \frac{u[m] \exp\frac{-i2\pi(m-1)(j-1)}{N_w}}{\sqrt{N_w}}, \quad j = 1, \cdots, N_w. \quad (21)$$

Thus, the average energies of the $i$th and $k$th sub-bands are

$$\text{AE}(SB_i) = \sum_{m=1}^{N_{wi}} \frac{|\tilde{u}[O_i+m]|^2}{N_{wi}} \triangleq \sum_{j=1}^{N_{\tilde{z}}} \frac{|\tilde{z}[j]|^2}{N_{\tilde{z}}} \triangleq \hat{M}_{\tilde{z}} \quad (22)$$

$$\text{AE}(SB_k) = \sum_{m=1}^{N_{wk}} \frac{|\tilde{u}[O_k+m]|^2}{N_{wk}} \triangleq \sum_{j=1}^{N_{\tilde{d}k}} \frac{|\tilde{d}_k[j]|^2}{N_{\tilde{d}k}}$$
$$\triangleq \hat{M}_{\tilde{d}k}, \quad \forall k \neq i$$

where $O_1 = 0$, $O_m = \lfloor (\sum_{j=1}^{m-1} B_j) T_w \rfloor, m \geq 2$, $N_{\tilde{z}} = N_{wi} = \lfloor T_w B_i \rfloor$, $N_{\tilde{d}k} = \lfloor T_w B_k \rfloor$, $\{\tilde{u}[O_i+m] = \tilde{z}[j]\}_{j=m=1}^{N_{\tilde{z}}}$ and $\{\tilde{u}[O_k+m] = \tilde{d}_k[j]\}_{j=m=1}^{N_{\tilde{d}k}}, \forall k \neq i$.

The $i$th sub-band contains noise only signal whereas, the $k$th sub-band contains signal plus noise. Thus, $\tilde{z}[j], \forall j$ can be modeled as i.i.d ZMCSCG random variables all with variance $\sigma^2$ and $\tilde{d}_k[j], \forall k \neq i, \forall j$ can be modeled as

$$\tilde{d}_k[j] = \tilde{s}_k[j] + \tilde{w}_k[j], \quad j=1,\cdots,N_{\tilde{d}k} \quad (23)$$

where $\tilde{w}_k[j], \forall k \neq i, \forall j$ are i.i.d ZMCSCG random variables all with variance $\sigma^2$ and $\tilde{s}_k[j], \forall k \neq i, \forall j$ are i.i.d zero mean random variables with $E\{|\tilde{s}_k[j]|^2\} = \gamma_{wk}\sigma^2$ and $\gamma_{wk}$ denotes the SNR of the $k$th sub-band. Without loss of generality, the samples $\tilde{s}_k[j]$, $\tilde{w}_k[j]$ and $\tilde{z}[j]$ are assumed to be independent.

For better explanation, define $\hat{R}_{wk}$ and $\tilde{R}_{wk}$ as

$$\hat{R}_{wk} \triangleq \frac{\hat{M}_{\tilde{d}k}}{\hat{M}_{\tilde{z}}} \quad (24)$$

$$\tilde{R}_{wk} \triangleq \frac{N_{\tilde{d}k} \to \infty \frac{1}{N_{\tilde{d}k}} \sum_{j=1}^{N_{\tilde{d}k}} |\tilde{d}[j]|^2}{N_{\tilde{z}} \to \infty \frac{1}{N_{\tilde{z}}} \sum_{j=1}^{N_{\tilde{z}}} |\tilde{z}[j]|^2} = \frac{M_{\tilde{d}k}}{M_{\tilde{z}}} = 1 + \gamma_{wk}$$

where $M_{\tilde{d}k} = N_{\tilde{d}k} \to \infty \frac{1}{N_{\tilde{d}k}} \sum_{j=1}^{N_{\tilde{d}k}} |\tilde{d}[j]|^2$ and $M_{\tilde{z}} = N_{\tilde{z}} \to \infty \frac{1}{N_{\tilde{z}}} \sum_{j=1}^{N_{\tilde{z}}} |\tilde{z}[j]|^2$. It follows

$$Pr[\text{AE}(SB_k) \geq \text{AE}(SB_i)] = Pr\left[\left(\frac{\hat{M}_{\tilde{d}k}}{\hat{M}_{\tilde{z}}} - 1\right) \geq 0\right], \forall k \neq i$$
$$= Pr\left[\sqrt{\tilde{N}_{dk}}\left(\hat{R}_{wk} - \tilde{R}_{wk}\right) \geq -\sqrt{\tilde{N}_{dk}}\gamma_{wk}\right] \geq \tilde{P}_{dwref}$$

where $\tilde{N}_{dk} = \frac{N_{\tilde{d}k} B_i}{(1+\gamma_{wk})^2 (B_i+B_k)} = \frac{T_w B_k B_i}{(1+\gamma_{wk})^2 (B_i+B_k)}, \forall k \neq i$.

By applying *Theorem 1*, it can be shown that

$$\sqrt{\tilde{N}_{dk}}\left(\hat{R}_{wk} - \tilde{R}_{wk}\right) \sim \mathcal{N}(0,1), \quad \forall k \neq i. \quad (25)$$

Thus,

$$Pr\left[\sqrt{\tilde{N}_{dk}}\left(\hat{R}_{wk} - \tilde{R}_{wk}\right) \geq -\sqrt{\tilde{N}_{dk}}\gamma_{wk}\right] =$$
$$\frac{1}{2}\text{erfc}\left(\frac{-\sqrt{\tilde{N}_{dk}}\gamma_{wk}}{\sqrt{2}}\right) \quad (26)$$
$$\Rightarrow T_w = 2\frac{B_i+B_k}{B_i B_k}\left(\left(1+\frac{1}{\gamma_{wk}}\right)\text{erfc}^{-1}(2\tilde{P}_{dwref})\right)^2$$
$$= \tau_k\left(\frac{1}{B_k} + \frac{1}{B_i}\right), \quad \forall k \neq i$$

where $\tau_k = 2\left(\left(1+\frac{1}{\gamma_{wk}}\right)\text{erfc}^{-1}(2\tilde{P}_{dwref})\right)^2$ and erfc(.) is the complementary error function [22]. As we can see from (26), $T_w$ depends on $B_i$, $B_k$ and $\gamma_{wk}$. Thus, to ensure (20), $T_w$ must be selected as

$$T_w = \tau\left(\frac{1}{\tilde{B}_1} + \frac{1}{\tilde{B}_2}\right) \quad (27)$$

where $\tau = \tau_k$ with $\gamma_{wk} = \gamma_w, \forall k \neq i$ and $\gamma_w$ is the target SNR of the cognitive device, $\tilde{B}_1$ and $\tilde{B}_2$ are the two smallest bandwidths of all of the sub-bands.

Now for the given $\tau$ (which is chosen by the cognitive device), what is the smallest possible $T_w$? As we can see from (27), for given $\tau$ and $2 \leq S_{sb} \leq S_{max}$, the minimum $T_w$ is achieved when $S_{sb} = 2$ and $B_i = B_k = \frac{B}{2}$. Hence,

$$T_{wmin} = \frac{4\tau}{B}. \quad (28)$$

Based on the analysis presented in this subsection and Section III-A, we describe the proposed edge and reference white sub-band detection algorithms in the following.

### C. Edge and Reference White Sub-band Detection Algorithms

Reliable edge detection plays an important role to guarantee the desirable performance for the proposed framework. Toward this end, we propose to impose the following constraints for detection and false alarm probabilities of the edge detection

$$P_{feL} = \tilde{P}_{feL} \quad (29)$$
$$P_{deL} \triangleq \min\{P_{deL1}, P_{deL2}\} \geq \tilde{P}_{deL} \quad (30)$$

where $P_{feL}$, $P_{deL1}$, and $P_{deL2}$ are given in (15), (18) and (19), respectively while $\tilde{P}_{feL}$ and $\tilde{P}_{deL}$ denote the target false alarm and detection probabilities. We can indeed control two parameters $T_{te}$ and $L$ to maintain these required constraints. It can be verified from (29) that for fixed targets $\tilde{P}_{feL}$ and $\tilde{P}_{deL}$, decreasing $T_{te}$ will increase the number of frames $L$ and vice verse. We propose to compute $L$ so as to attain $P_{feL} = \tilde{P}_{feL}$ and $T_{te} = T_{wmin}$, where $T_{wmin}$ is the minimum received signal duration of each frame which is given in (28). In this paper, we set $\tilde{P}_{feL} = 0.001$ and $\tilde{P}_{deL} = 0.999$ to avoid error propagation. Note that the sensing time required to maximize the total throughput $T_o$, which will be described in Section V, is typically larger than $T_{wmin}$. Therefore, the computed value of $L$ is guaranteed to maintain the constraints under the sensing time $T_o$.

The proposed edge and reference white sub-band detection algorithms are summarized as follows:

**Algorithm I**: Edge and Reference white sub-band detection.
**Inputs**: $B$, $B_{min}$, $S_{max}$, $\tilde{P}_{dwref}$, $\tilde{P}_{feL}$, $\tilde{P}_{deL}$, $\gamma_e = \gamma_w$.

1) Compute $T_{wmin}$ using (28) and set $T_{et}$ of (3) as $T_{wmin}$.
2) Using $T_{wmin}$, compute $N_e^\star = \lfloor T_{wmin} B \rfloor$ and $N_{eh}^\star = \lfloor \frac{T_{wmin} B_{min}}{2} \rfloor$.
3) Using $N_{eh}^\star$, compute $\lambda_e^\star$ and $L$ ensuring $P_{feL} = \tilde{P}_{feL}$ of (29) and $P_{deL} \geq \tilde{P}_{deL}$ of (30).
   **Obj 1**: Edge detection
   Initialize f=0 and $\mathbf{q} = \mathbf{0}^{N_e^\star \times 1}$.
   **Repeat:** f=f+1.
   a) In frame $f$, take $N_e^\star$ samples and set them as $\mathbf{x}_f$.

b) Using $\mathbf{x}_f$ and $N_{eh}^\star$, compute $\tilde{R}_e^2$ of (14) and set the resulting values as $\mathbf{q}_f$.
c) Update $\mathbf{q} = \mathbf{q} + [\mathbf{0}^{2N_{eh} \times 1}; \mathbf{q}_f; \mathbf{0}^{2N_{eh} \times 1}]$.
**Until:** f=L-1.
d) Set False=0.
**Repeat**
  ⋄ Get j= $\arg\max \mathbf{q}$.
  **if** $q[j] \geq \lambda_e^\star$ **then**
    ⋄ Decide the $j$th frequency as edge.
    ⋄ Reduce the size of $\mathbf{q}$ by excluding the frequencies $[j - B_{min} : j + B_{min}]$.
  **else**
    False=1
  **end if**
**Until**: False=1.
e) From these edges, determine the sub-bands $SB_1, SB_2, \cdots, SB_{S_{sb}}$.

**Obj 2**: Reference white sub-band isolation
a) From $SB_1, SB_2, \cdots, SB_{S_{sb}}$, compute $T_w$ using (27).
b) From the received $T_w B$ samples, compute the average energies of $SB_1, SB_2, \cdots, SB_{S_{sb}}$.
c) The sub-band that has the minimum average energy is set as the reference white sub-band $SB_i$.

**Remark**: When all of the sub-bands contain noise only signal, the number of edges could be 0. In this case, we set $S_{sb} = 2$ where the bandwidth of each sub-band is $\frac{B}{2}$. As will be clear later, in each frame the cognitive device will receive samples of duration $T_o > T_{wmin}$. Hence, in practice $P_{deL} \approx 1$ can be achieved by utilizing all the available samples of the previous $L-1$ frames.

## IV. GENERALIZED ENERGY DETECTION

This section discusses the proposed GED to label each of the target sub-bands $SB_k, k = 1, \cdots i-1, i+1, \cdots S_{sb}$ as white or non-white (i.e., **Obj 3**). Here we assume that the cognitive device employs $T_{ts} = T_w + T_a$, where $T_w$ is the time to detect the $i$th reference white sub-band (i.e., (27)) and $T_a$ is the additional sensing time to ensure the target $\tilde{P}_d$.

In $T_{ts}$ interval, the cognitive device will have $N_s = \lfloor T_{ts} B \rfloor$ samples. By denoting these samples as $\bar{u}[j]$, the DFT of $\bar{u}[j], \forall j$ can be expressed as

$$\tilde{\bar{u}}[j] = \sum_{m=1}^{N_s} \frac{\bar{u}[m] \exp^{\frac{-i2\pi(m-1)(j-1)}{N_a}}}{\sqrt{N_s}}, \quad j = 1, \cdots, N_s. \quad (31)$$

Like in (22), the average energies of the $i$th reference white sub-band and the $k$th sub-band can be expressed as

$$\text{AE}(SB_i) = \frac{1}{N_z} \sum_{j=1}^{N_z} |z[j]|^2 \triangleq \hat{M}_z$$

$$\text{AE}(SB_k) = \frac{1}{N_{dk}} \sum_{j=1}^{N_{dk}} |d_k[j]|^2 \triangleq \hat{M}_{dk}, \quad \forall k \neq i \quad (32)$$

where $N_{dk} = \lfloor T_{ts} B_k \rfloor$ and $N_z = \lfloor T_{ts} B_i \rfloor$. As the $i$th reference white sub-band contains noise only signal, $z[j], \forall j$ can be modeled as i.i.d ZMCSCG random variables all with variance $\sigma^2$ whereas, the $k$th sub-band contains either noise only or signal plus noise. Thus, $d_k[j], \forall k \neq i, \forall j$ can be modeled as

$$d_k[j] = \begin{cases} \tilde{\tilde{s}}_k[j] + \tilde{\tilde{w}}_k[j], & \text{Under } H_{1k} \\ \tilde{\tilde{w}}_k[j], & j = 1, \cdots, N_{dk} \text{ Under } H_{0k} \end{cases} \quad (33)$$

where $\tilde{\tilde{w}}_k[j], \forall j$ are i.i.d ZMCSCG random variables all with variance $\sigma^2$, $\tilde{\tilde{s}}_k[j], \forall j$ are i.i.d zero mean random variables with $\text{E}\{|\tilde{\tilde{s}}_k[j]|^2\} = \gamma_k \sigma^2$ and $\gamma_k$ denotes the SNR of the $k$th sub-band under $H_{1k}$ hypothesis.

To detect the $k$th sub-band, we propose the following test statistics

$$R_k = \sqrt{\frac{N_{dk} \beta_k}{\beta_k + 1}} \left( \frac{\hat{M}_{dk}}{\hat{M}_z} - 1 \right), \quad \forall k \neq i \quad (34)$$

where $\beta_k = \frac{N_z}{N_{dk}} = \frac{B_i}{B_k}$. By applying *Theorem 1*, it can be shown that

$$R_k \sim \mathcal{N}(0, 1), \quad \text{Under } H_{0k}$$
$$R_k \sim \mathcal{N}(\mu_k, \tilde{\sigma}_{H_{1k}}^2), \quad \text{Under } H_{1k}$$

where $\mu_k = \sqrt{\frac{N_{dk} \beta_k}{\beta_k + 1}} \gamma_k$ and $\tilde{\sigma}_{H_{1k}} = 1 + \gamma_k$. The probability of detection and false alarm of the test statistics (34) are

$$P_{fk}(\lambda_k) = Pr(R_k > \lambda_k | H_{0k}) = \frac{1}{2} \text{erfc} \left( \frac{\lambda_k}{\sqrt{2}} \right) \quad (35)$$

$$P_{dk}(\lambda_k) = Pr(R_k > \lambda_k | H_{1k}) = \frac{1}{2} \text{erfc} \left( \frac{\lambda_k - \mu_k}{\sqrt{2} \tilde{\sigma}_{H_{1k}}} \right) \quad (36)$$

where $\lambda_k$ is the threshold. As we can see from (36), for the given $\gamma_k > 0$ and $\lambda_k$, increasing $N_{dk}$ increases $P_{dk}$. This is due to the fact that $\text{erfc}(.)$ is a decreasing function. Thus, the proposed detection algorithm is consistent and does not suffer from any SNR wall (i.e., for any given $P_{fk} > 0$ and $\gamma_k > 0$, $P_{dk} \to 1$ as $N_{dk} \to \infty$). One can also notice that the detector (34) is not very sensitive to small to medium interference signal. This is because, the ratio $\frac{\hat{M}_{dk}}{\hat{M}_z}$ will not be changed significantly in the presence of small to medium interference signals. Thus, the proposed detector is robust against small to medium ACI which will occur frequently in practice.

In the following, we address the relation between the detector (34) and the CED. As can be seen from (34), when $\beta_k \to \infty$, $\hat{M}_z$ becomes the true noise variance and the test statistics (34) will be

$$R_{k\beta_k \to \infty} = \sqrt{N_{dk}} \left( \frac{\hat{M}_{dk}}{\hat{M}_z} - 1 \right) = \sqrt{N_{dk}} \left( \frac{\hat{M}_{dk}}{\sigma^2} - 1 \right). \quad (37)$$

Indeed this is shifted and scaled version of the CED which is optimal. The numerator term of (34) is computed from the power of the desired sub-band and the denominator term is computed from the power of the reference white sub-band. As the reference sub-band contains noise information only, the denominator term of (34) can be regarded as the estimate of the noise power which is the true noise variance when $\beta_k \to \infty$ (i.e., as in (37)). For this reason, we have termed the test statistics (34) as a GED. However, the key difference between our GED and the CED is that the latter suffers from SNR wall whereas, the former does not.

Next we examine the following interesting question. For the given $\beta_k$, how much is the performance loss of (34) compared to that of the CED? As can be seen from (35), $P_{fk}$ does not depend on $\beta_k$. Thus, the test statistics (34) and (37) will employ the same threshold $\lambda_k$ to ensure a certain $P_{fk}$. This threshold is given by

$$\lambda_k^\star = \sqrt{2}\text{erfc}^{-1}(2P_{fk}), \quad \forall k \neq i.$$

Thus, the detection performance loss is given as

$$\eta_k = 1 - \frac{P_{dk}(R_k)}{P_{dk}(R_{k\beta_k \to \infty})} = 1 - \frac{\text{erfc}\left(\frac{\lambda_k^\star - \sqrt{\frac{N_{dk}\beta_k}{\beta_k+1}}\gamma_k}{\sqrt{2}(1+\gamma_k)}\right)}{\text{erfc}\left(\frac{\lambda_k^\star - \sqrt{N_{dk}}\gamma_k}{\sqrt{2}(1+\gamma_k)}\right)}.$$

From these explanations, the following key points can be highlighted:

1) If the noise variance is estimated from finite sub-band, the theoretical thresholds of the CED can not be applied directly.
2) When the bandwidth of the white sub-band is very small (i.e., $\beta_k$ is very small), the threshold value to ensure a certain $P_{fk}(P_{dk})$ of (34) is significantly different from that of the CED.
3) Increasing $\frac{\beta_k}{\beta_k+1}$ increases the detection performance of (34). Hence, the detection performance of the proposed GED is upper bounded by that of the CED.

The paper aims to accomplish four objectives. In the second objective, a reference white sub-band is selected from all available sub-bands. As can be seen from (34), the power of this reference white sub-band is used as the estimate of the noise variance. When the reference white sub-band is chosen incorrectly, we will have $\hat{M}_z > \sigma^2$. And in such a case, the theoretical $P_d$ and $P_f$ can not be ensured. Thus, correct identification of the reference white sub-band is crucial to maintain the desired detection and false alarm probabilities.

## V. Sensing Time Optimization

In this section, we compute the optimal $T_{ts}$ of (31) to maximize the total throughput of the CR network (i.e., **Obj 4**). The CR network performs transmission when the GED (34) declares a given target sub-band as white. The proposed generalized energy detection algorithm has a certain missed detection (i.e., the GED (34) may declare a non-white sub-band as white). Thus, in the $k$th sub-band, the CR network can have the following two SNRs [14]:

$$\gamma_c|H_{0k} = \gamma_c, \qquad \text{Correct sensing decision}$$
$$\gamma_c|H_{1k} = \frac{\gamma_c}{1+\gamma_{pk}}, \quad \text{Incorrect sensing decision} \quad (38)$$

where $\gamma_c$ is the SNR of the CR network and $\gamma_{pk}, \forall k \neq i$ are the SNR of the primary signal experienced at the receiver of the CR network[7]. If we denote the probability of the occurrences of $H_{0k}$ and $H_{1k}$ by $P(H_{0k})$ and $P(H_{1k})$[8], respectively, in the transmission time duration $T_f - T_o$ (recall Fig. 2), we will achieve the following two throughputs:

$$Th_{H_{0k}} = \frac{T_f - T_o}{T_f} R_{0k} P(H_{0k})(1 - P_{fk}(\tilde{\lambda}_k, T_o))$$
$$Th_{H_{1k}} = \frac{T_f - T_o}{T_f} R_{1k} P(H_{1k})(1 - P_{dk}(\tilde{\lambda}_k, T_o))$$

where $R_{0k} = \log_2(1+\gamma_c|H_{0k})$ and $R_{1k} = \log_2(1+\gamma_c|H_{1k})$[9]. Our objective will now be to get the optimal $T_o$ for maximizing the sum of the throughputs of all the target sub-bands under the constraint that the primary network(s) (i.e., each target sub-band) is sufficiently protected. This problem is mathematically formulated as

$$\max_{T_o} \sum_{k=1, k\neq i}^{S_{sb}} (Th_{H_{0k}} + Th_{H_{1k}})$$
$$\text{s.t } P_{dk}(\tilde{\lambda}_k, T_o) \geq \tilde{P}_d, \quad \forall k \neq i \quad (39)$$

where $\tilde{P}_d$ is the target detection probability of each sub-band. As can be seen from this expression, $P_{dk}$ depends on $\tilde{\lambda}_k$ and $T_o$. Furthermore, for the given $T_o$, the optimal $\tilde{\lambda}_k$ of the above problem can be obtained by setting

$$P_{dk}(\tilde{\lambda}_k, T_o) = \tilde{P}_d, \quad \forall k \neq i. \quad (40)$$

From (36), we will have

$$P_{dk}(\tilde{\lambda}_k) = \frac{1}{2}\text{erfc}\left(\frac{\tilde{\lambda}_k - \mu_k}{\sqrt{2}\tilde{\sigma}_{H_{1k}}}\right). \quad (41)$$

By combining (40) and (41), the optimal $\tilde{\lambda}_k$ becomes

$$\tilde{\lambda}_k^\star = a_k\sqrt{T_o} + b_k, \quad \forall k \neq i \quad (42)$$

where $a_k = \sqrt{\frac{\beta_k B_k}{\beta_k+1}}\gamma_k$, $b_k = \sqrt{2}\tilde{\sigma}_{H_{1k}}\text{erfc}^{-1}(2\tilde{P}_d)$. Substituting $\tilde{\lambda}_k^\star, \forall k \neq i$ into (35) and after some straightforward steps, problem (39) can be reformulated as

$$\max_{T_o} \frac{T_f - T_o}{T_f} \sum_{k=1, k\neq i}^{S_{sb}} \left(\psi_k \text{erf}\left(\frac{a_k\sqrt{T_o} + b_k}{\sqrt{2}}\right) + \tilde{\psi}_k\right)$$
$$\triangleq \tilde{f}(T_o) \quad (43)$$

where $\psi_k = 0.5 P(H_{0k})\log_2(1+\gamma_c|H_{0k})$ and $\tilde{\psi}_k = \psi_k + P(H_{1k})\log_2(1+\gamma_c|H_{1k})(1-\tilde{P}_d)$ are constants, and $\text{erf}(.) = 1 - \text{erfc}(.)$.

The optimal $T_o$ of this problem can be obtained numerically by simple bisection search method [23], [24] (see Appendix B for the details).

Similar optimization problem has been examined in [14] for CED. However, the authors of [14] examine their problem by ignoring $Th_{H_1}$ (see equation (21) of [14]). The proposed generalized energy detection and sensing time optimization algorithms are summarized as follows.

**Algorithm II**: Generalized Energy Detection and Sensing Time Optimization
**Inputs**: $B$, $B_i$ and $B_k, \forall k \neq i$ and $\gamma_k$.

---

[7]In practice, we do not have any information about $\gamma_{pk}, \forall k \neq i$. Due to this fact, we employ $\gamma_{pk} = \gamma_k, \forall k \neq i$.

[8]These probabilities can be computed by employing the decision statistics of the previously sensed frames (i.e., the decision statistics of (34)).

[9]Here we assume that the CR network transmits a Gaussian signal and the channel between the CR transmitter and receiver is assumed to be AWGN.

1) **Obj 4**: Optimal sensing time computation
   a) From $B_i$ and $B_k, \forall k \neq i$, determine $T_o^\star$ that maximize the throughput by solving (43).
   b) Get the remaining samples $(T_o^\star - T_w)B$[10] and denote the total received samples as $\{\bar{u}[j]\}_{j=1}^{N_s}$ of (31), where $N_s = \lfloor T_{ts}B \rfloor$ with $T_{ts} = T_o^\star$.
2) **Obj 3**: Generalized energy detection
   a) Using these $N_s$ samples $\{\bar{u}[j]\}_{j=1}^{N_s}$ and the reference white sub-band $SB_i$, compute $R_k$ with (34), $\forall k \neq i$.
   b) Compute $\lambda_k^\star$ using (42) and
      **if** $R_k < \lambda_k^\star$ **then**
         Label the $k$th sub-band as white.
      **else**
         Label the $k$th sub-band as non-white.
      **end if**
3) **Transmission**: Use the white space sub-bands for transmitting information over the CR network in the remaining $T_f - T_o$ seconds. In this time duration, the edges of the next frame can also be determined.

In the considered CR system, it is assumed that the noise power levels across the frequency region of interest are the same. Furthermore, the channel between each primary transmitter and cognitive device is assumed to be AWGN. Thus, the primary transmitters have the same spatial statistical behavior. However, the positions of primary transmitters may not be relatively close to each other, and the signal levels of primary transmitters may not be necessarily similar. From this explanation, we can understand that the current paper examines its problems for similar noise and spatial statistical behaviors across the spectral region of interest. Therefore, although the detection approaches of the current paper can still be applied for any fading channel environments, the analysis of the proposed algorithms for non uniform noise power and mixed statistical channel environments (e.g., a mix of AWGN and Rayleigh fading channels) is still an open research topic. Note also that the proposed algorithms of this paper can be extended straightforwardly to multiple antenna cognitive device scenario. The details are omitted for conciseness.

## VI. NUMERICAL EXAMPLES

In this section we provide some of the parameters of this paper. Currently, the FCC has proposed TV bands for CR network application [14]. The bandwidth of each TV band is 6MHz and the target $\tilde{P}_d$ is $\tilde{P}_d = 0.9$ at $\gamma = -20$dB. Suppose that we have a wide-band cognitive device with bandwidth 60MHz (i.e., for 10TV bands), $\tilde{P}_{deL} = 0.999$, $\tilde{P}_{feL} = 0.001$, $\tilde{P}_{wdref} = 0.999$, $\gamma_c = 20$dB, $\gamma_e = \gamma_w = \gamma_k = -20$dB, $\tilde{P}_{dk} = 0.9$, $P(H_{0k}) = 0.8$ and $P(H_{1k}) = 0.2, \forall k \neq i$. For these settings, according to assumption 7 of Section II, the minimum desired bandwidth will be $B_{min} = 6$MHz.

1) From (28), $T_{wmin}$ becomes $T_{wmin} = 6.5$ms.
2) By using (29) and (30), we will get $L = 55$.

---

[10]This is because $T_w B$ samples are already taken during the reference white sub-band isolation phase.

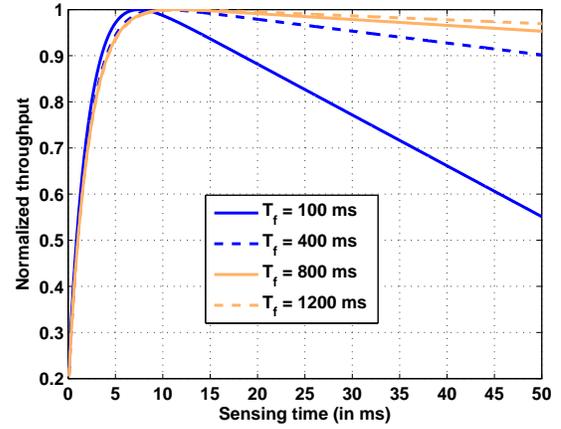

(a)

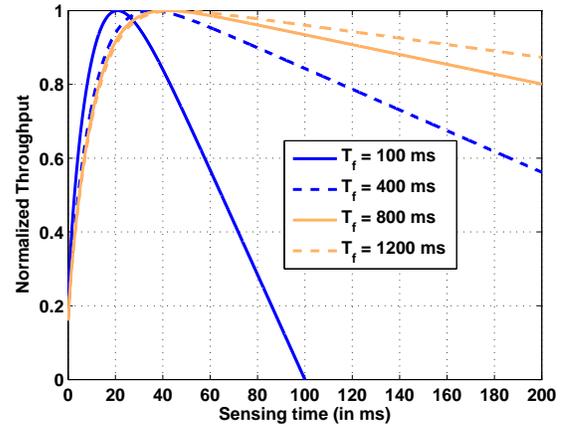

(b)

Fig. 4. Normalized $\tilde{f}(T_o)$ of (43) for $\gamma_c = 20$dB, $\gamma_e = \gamma_w = \gamma_k = -20$dB and different frame duration $T_f$. (a) when $S_{sb} = 2$, (b) when $S_{sb} = S_{max} = 10$.

As can be seen from (43), the optimal $T_o$ depends on the reference white sub-band $SB_i$ and the sub-bands $SB_k, \forall k \neq i$. As these sub-bands are not necessarily the same for all frames, the optimal $T_o$ may vary from one frame to the other. However, the minimum and maximum $T_o$ can be obtained by examining two extreme cases, where the first case is when $S_{sb} = 2$ with $B_1 = B_2 = \frac{B}{S_{sb}} = \frac{B}{2}$ (this corresponds to the minimum $T_o$) and the second case is when $S_{sb} = S_{max}$ with $\{B_j = \frac{B}{S_{sb}} = \frac{B}{S_{max}}\}_{j=1}^{S_{max}}$ (which corresponds to the maximum $T_o$).

Fig. 4 shows $\tilde{f}(T_o)$ of (43) for these two cases. As can be seen from this figure, lower sensing time is required when $S_{sb} = 2$ and higher sensing time is required when $S_{sb} = S_{max} = 10$ which is expected. Furthermore, in both cases, when the frame duration increases, the optimal sensing time also increases. However, this increment is not linear. For example, from Fig. 4.(b), we can see that the optimal sensing times with $T_f = 100$ms and $T_f = 1200$ms (i.e., 12 times increment) are 20ms and 45ms (i.e., 2.25 times increment), respectively. Thus, for practical application it is desirable to choose the maximum possible frame duration. For 802.22 system, we suggest to set $T_f = 2$ seconds. From this figure, we can also notice that $T_{wmin} \leq T_o$ which confirms that $T_o > T_{wmin}$ for these practical settings.

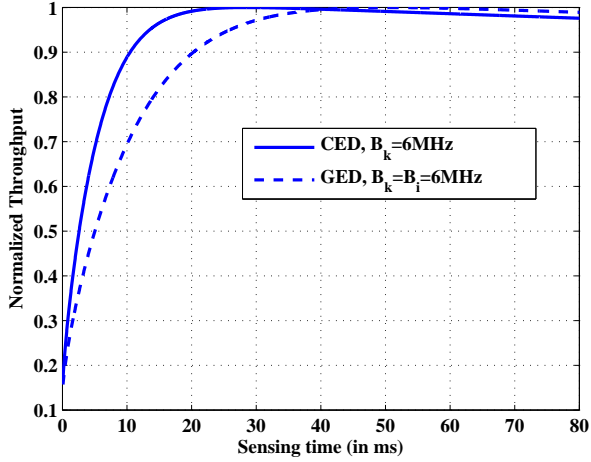

Fig. 5. Comparison of the sensing time of the CED and GED with $B_i = B_k = 6$MHz, $\forall i \neq k$, $\gamma_c = 20$dB, $\gamma_e = \gamma_w = \gamma_k = -20$dB and frame duration $T_f = 2$s, .

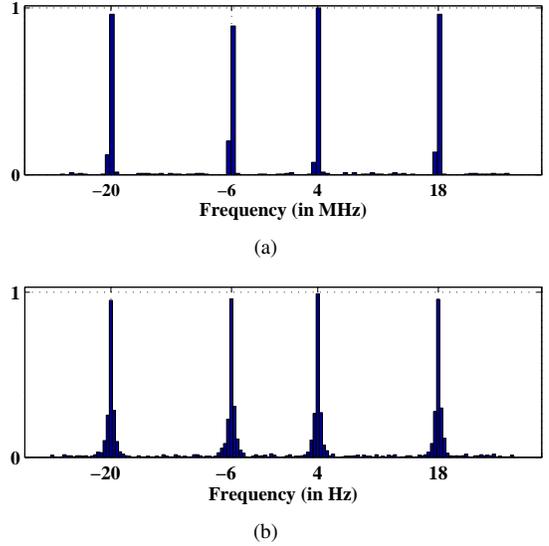

Fig. 6. Normalized histogram of the edge detection algorithm in AWGN channel: (a) $\gamma_e = -20$dB, (b) $\gamma_e = -22$dB.

Next we compare the sensing time of the GED of (34) and that of the CED for the target sub-band with $T_f = 2$s and $S_{sb} = S_{max} = 10$ which is shown in Fig. 5. As can be seen from this figure, the CED ($T_{oCED} = 28.5$ms) requires less sensing time compared to that of the proposed GED ($T_{oGED} = 50.6$ms) (i.e., $T_{oGED} \approx 1.78 T_{oCED}$). This is expected since the CED assumes perfect noise variance (i.e., $\beta_k \to \infty$). This result validates that the maximum throughput is achieved when the noise variance is known perfectly.

Throughout the analysis of the current paper, we employ Nyquist sampling, and the industries fastest ADCs can not support more than 250Msps (e.g., ANALOG DEVICES ADC) which corresponds to around $B = 250$MHz. As we can see from (4), the proposed algorithms can be applied when each of the samples of the received signal is independent. In fact, the independence behavior of the received samples can be maintained by uniform sub-Nyquist sampling. This shows that the proposed algorithms can still be applied for uniform sub-Nyquist sampling. However, as this sampling approach reduces the number of samples, more sensing time is required to ensure the same performance as that of the full Nyquist rate. For example, when the sampling rate is halved, the sensing time will be doubled (compared to the Nyquist rate). From this explanation, one can understand that the algorithms considered in the current paper are not efficient when sub-Nyquist sampling is applied. And extending the proposed algorithms efficiently for sub-Nyquist sampling approach is an open research topic (see [25] and [26] for a survey of different spectrum sensing and sampling approaches for wideband CR networks).

The spectrum utilization of primary networks at a given location and time is extremely low [16]. This validates that the bandwidth $B = 60$MHz (i.e.,10 TV bands) which incorporates more than one sub-bands will have at least one white sub-band with very high probability. From this explanation, one can understand that the analysis of the current paper are valid for moderate speed ADCs (i.e., 60Msps).

## VII. SIMULATION RESULTS

This section provides simulation results which are obtained by averaging 20000 experiments. For the simulation, we consider a wide-band signal of bandwidth $B = 60$MHz (i.e., from $-\frac{B}{2}$ to $\frac{B}{2}$) and $B_{min} = 6$MHz (i.e., $S_{max} = 10$). The true number of sub-bands are set to $S_{sb} = 5 < S_{max}$ with $SB_1 = [-30 : -20]$, $SB_2 = [-20 : -6]$, $SB_3 = [-6 : 4]$, $SB_4 = [4 : 18]$ and $SB_5 = [18 : 30]$MHz. When a sub-band contains signal plus noise, the signal is taken from the quadrature phase shift keying (QPSK) constellation with $\sigma_s^2 = 1$mw. Furthermore, the channel between the primary transmitter and the cognitive device is AWGN (i.e., in each sub-band, the channel between the transmitter and the cognitive device is AWGN). The SNR is defined as $SNR \triangleq \frac{\sigma_s^2}{\sigma^2}$.

### A. Performance of the Edge Detector

In this simulation, we verify the theoretical edge detection algorithm with simulation. In each experiment (frame), we utilize either of the following two scenarios.

1) The sub-bands $SB_2$ and $SB_4$ contain signal plus noise with SNR $\gamma_e$ whereas, the sub-bands $SB_1$, $SB_3$ and $SB_5$ contain noise only signal.
2) The sub-bands $SB_1$, $SB_3$ and $SB_5$ contain signal plus noise with SNR $\gamma_e$ whereas, the sub-bands $SB_2$ and $SB_4$ contain noise only signal.

It can be noticed that the number of edges in either of these scenarios is 4. However, the PSD of each sub-band is different from one scenario to the other. The duration of each experiment is set to $T_{te} = T_{wmin} = \frac{4\tau}{B} = 6.5$ms and $L = 55$. Fig. 6 shows the normalized histogram of the edge detection algorithm for $\gamma_e = -20$dB and $-22$dB. As can be seen from this figure, the proposed edge detector reliably detect all of the edges (i.e., $[-20, -6, 4, 18]$MHz) almost accurately (which means $P_{deL} \approx 1$) when SNR$=-20$dB. Furthermore, the accuracy of the proposed edge detection algorithm worsens when the SNR decreases which is expected.

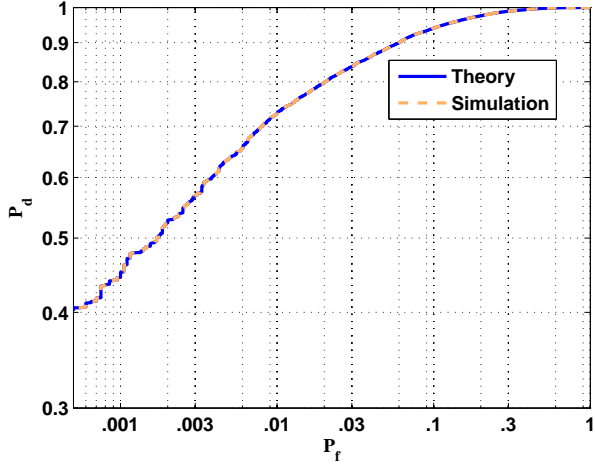

Fig. 7. Comparison of theoretical and simulated $P_f$ versus $P_d$ of GED in AWGN channel at SNR$= -20$dB.

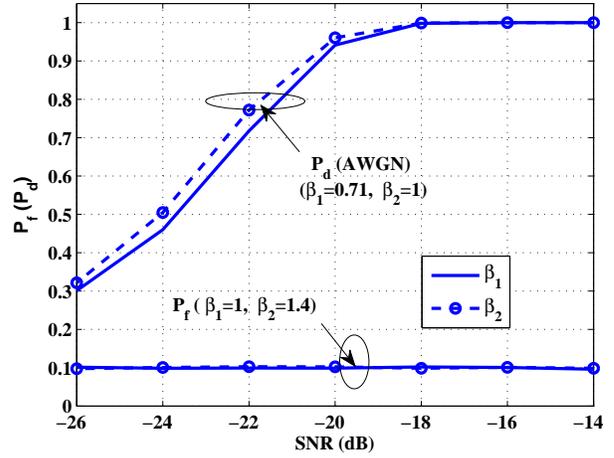

Fig. 8. Performance of the proposed GED under noise variance uncertainty.

TABLE I
PERFORMANCE OF THE REFERENCE WHITE SUB-BAND DETECTOR

| $\gamma_w$(dB) | -14 | -16 | -18 | -20 | -22 |
|---|---|---|---|---|---|
| $T_w$(ms) | 1.3 | 3.2 | 7.8 | 19.5 | 48.6 |
| $P_{dwref}$ | 0.9991 | 0.999 | 0.99901 | 0.9989 | 0.999 |

### B. Performance of the Reference White Sub-band Detector

In this simulation, we verify the theoretical results of the reference white sub-band detector with simulation. To this end, we consider the two scenarios explained in the previous subsection. By applying (27) with $\tilde{P}_{dwref} = 0.999$, we compute the required $T_w$ for different SNR values (i.e., $\gamma_w$). And, with these $\gamma_w$ and the corresponding $T_w$, the simulated $P_{dwref}$ is shown in Table I. As can be seen from this table, $T_w$ increases as $\gamma_w$ decreases and $P_{dwref} \approx \tilde{P}_{dwref}$ for all $\gamma_w$.

### C. Verification of $P_f$ versus $P_d$ Expressions of GED (34)

In this subsection, we verify the theoretical $P_{fk}$ and $P_{dk}$ expressions of the GED (34) by computer simulations. To this end, we use sub-band $i = 3$ ($SB_3$) as a reference white sub-band. For the $P_f$ expression, we assume that the first sub-band (i.e., $k = 1$) contains noise only signal. And for the $P_d$ expression, we assume that the second sub-band (i.e., $k = 2$) contains signal plus noise with SNR$=-20$dB. The $T_{ts}$ of (31) is set to $T_{ts} = 13$ms. Under these settings, Fig. 7 shows the comparison of the theoretical and simulated $P_f$ versus $P_d$ results. From this figure, we can see that the theoretical $P_f(P_d)$ matches exactly that of the simulated one.

### D. Effect of $\beta_k$ and Noise Variance Uncertainty on the $P_{fk}$ and $P_{dk}$ of GED

In this subsection, we examine the effect of $\beta_k$ and noise variance uncertainty on the $P_{fk}$ and $P_{dk}$ of the proposed GED. In an uncertain noise variance scenario, the true noise variance can be modeled as a bounded interval of $[\frac{1}{\epsilon}\sigma^2 \;\; \epsilon\sigma^2]$ for some $\epsilon = 10^{\Delta\sigma^2/10} > 1$, where the uncertainty $\Delta\sigma^2$ is expressed in dB [4]. We assume that this bound follows a uniform distribution, i.e., $\mathcal{U}[\frac{1}{\epsilon}\sigma^2 \;\; \epsilon\sigma^2]$. The noise variance is the same for one experiment (since it has a short duration) and follows a uniform distribution during several experiments. For this simulation, we set $\Delta\sigma^2 = 2$dB, $T_{ts} = 13$ms and the target $P_f$ is set to $P_f \leq 0.1$.

For better exposition of this simulation, we consider two reference white sub-bands $i = [3, 4]$. For the $P_f$ expression, we assume that the first sub-band (i.e., $k = 1$) contains noise only signal. And for the $P_d$ expression, we assume that the second sub-band (i.e., $k = 2$) contains signal plus noise with different SNRs. Hence, for the noise only sub-band (i.e., $k = 1$), we will have $\beta_1 = \frac{B_3}{B_1} = 1$ and $\beta_2 = \frac{B_4}{B_1} = 1.4$. And for the signal plus noise sub-band (i.e., $k = 2$), we will have $\beta_1 = \frac{B_3}{B_2} = 0.7143$ and $\beta_2 = \frac{B_4}{B_2} = 1$. Fig. 8 shows the achieved $P_f$ and $P_d$ values for these $\beta$ values. From this figure we can understand that the target $P_f \leq 1$ is maintained for both $\beta$ values of $SB_1$. Thus, the $P_f$ of the detector (34) does not depend on the value of $\beta$ which is inline with the theoretical result. Furthermore, in the case of signal plus noise scenario (i.e., $SB_2$), increasing $\frac{\beta}{\beta+1}$ (or SNR) increases the detection performance of the detector (34).

### E. Comparison of the Proposed GED and the detector of [11] for Pulse Shaped Signals

Recently new linear combination approach signal detection algorithm is proposed for pulse shaped transmitted signals with known rolloff factor in [11]. The detector of [11] is robust against noise variance uncertainty and small to medium ACI, and it outperforms CED and EVD-based signal detectors. Furthermore, the detector of [11] is already implemented using universal software radio peripheral (USRP) in [27] and has shown consistent result with the theory. Due to this reason, we compare the proposed GED (34) with the detector of [11] for pulse shaped signals with known rolloff factor (i.e., one band with known rolloff). To this end, we consider that the transmitted signal is pulse shaped by a square root raised cosine filter (SRRCF) with period $T_s$ and a certain rolloff factor (i.e., the total bandwidth of the transmitted signal is $\frac{(1+\text{rolloff})}{T_s}$Hz).

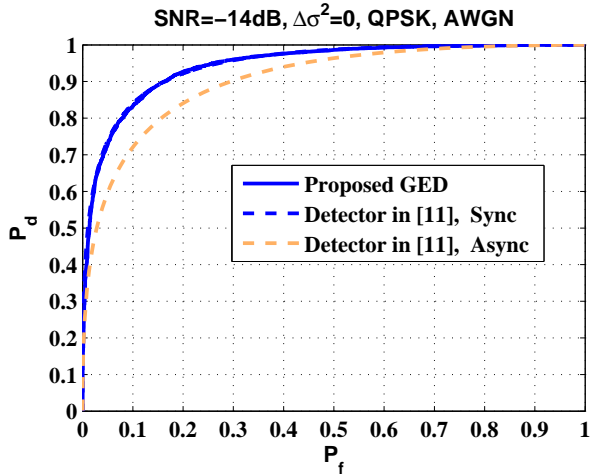

Fig. 9. Comparison of the proposed GED and the detector in [11] for pulse shaped transmitted signals with $T_{ts} = 4.55$ms.

For the comparison, we consider the same scenario as in Fig. 3 of [11] (i.e., $T_s = \frac{1}{6}10^{-6}$s, rolloff $= 0.2$, $T_{ts} = 4.55$ms and AWGN channel). From fundamental wireless communication, it is known that the rolloff frequency regions of any pulse shaped signal are highly dominated by the noise (i.e., it contains almost noise only signal). Therefore, one can interpret that any pulse shaped transmitted signal has two sub-bands, where the first sub-band of bandwidth $\frac{1}{T_s}$Hz contains signal plus noise (under $H_1$ hypothesis) and the second sub-band of bandwidth $\frac{\text{rolloff}}{T_s}$Hz contains noise only signal. Hence, for our generalized energy detection algorithm (34), the former sub-band can be considered as the target sub-band (i.e., sub-band $SB_k=[-3:3]$MHz) and the latter sub-band can be considered as the white sub-band (i.e., sub-band $SB_i=[-3.6:-3]$MHz and $[3:3.6]$MHz). Due to this reason, for the set up of Fig. 3 of [11], (34) utilizes, $B_i = 1.2$MHz, $B_k = 6$MHz and $\beta_k = \frac{B_i}{B_k} = \text{rolloff} = 0.2$.

Fig. 9 shows the performances of the proposed GED and that of the detector in [11]. From this figure, we can see that the current algorithm achieves the same performance as that of [11] when the cognitive device is perfectly synchronized with the primary transmitter (i.e., Sync) which almost never happen in practice. However, for the practically relevant asynchronous scenario (i.e., Async), the current GED (which do not assume any synchronization) outperforms the detector in [11]. From this result, we can also conclude that the proposed GED outperforms CED and EVD-based detectors under noise variance uncertainty. On the other hand, the proposed GED requires neither the pulse shaping filter of the transmitted signal nor oversampling of the received signal.

**Complexity analysis**: In the following, we examine the complexities of the EVD-based detection algorithm of [8], Max-Min SNR energy based algorithm of [11], CED and proposed GED algorithms. For convenience, we compare the complexities of these algorithms for a pulse shaped transmitted signal with rolloff factor $0.2$. For this signal, let us denote $\tilde{N}$ as the number of samples obtained by employing Nyquist sampling. As mentioned in the introduction section, the approaches of [8] and [11] apply oversampling factor beyond Nyquist rate which is assumed to be $\tilde{L}$. Under such assumptions, the complexities of these algorithms are summarized in Table II. In this table, one flop is defined as one complex multiplication and addition. The EVD-based algorithm of [8] utilizes $0.8\tilde{N}\tilde{S}$ flops which is used to compute the covariance matrix (21) of [8], where $\tilde{S} \geq \tilde{L}$ is a smoothing factor. And the complexity $O(\tilde{S}^3)$ is required to compute the maximum and minimum eigenvalues of this covariance matrix, and the pre-whitening matrix $\mathbf{Q}^{-1}$ in (48) of [8]. For the algorithm of [11], $1.6\tilde{N}\tilde{L}^2$ flops are required, where $0.8\tilde{N}\tilde{L}^2$ flops are used to compute the average power corresponding to the maximum (minimum) SNR samples. For the CED and proposed GED algorithms, $\tilde{N}$ point FFT is required to convert the time domain samples to frequency domain and $\tilde{N}$ flops are required to compute the average power of the considered band. The complexity of the proposed GED algorithm and that of the CED are the same, and by employing Radix-2 algorithm, the FFT operation can be performed with $O(\tilde{N}\log(\tilde{N}))$ complexity. As can be seen from this table, the complexities of these algorithms are close to each other. However, the proposed GED algorithm (also CED) does not require over sampling which is desirable as it can be implemented with low cost ADC.

We would like to mention here that the computational complexities of the edge detection and reference white sub-band detection algorithms are similar to that of the GED algorithm. The detailed complexity analysis of these algorithms is omitted for conciseness.

TABLE II
COMPLEXITIES OF DIFFERENT ALGORITHMS

| Algorithm Type | Complexity |
| --- | --- |
| EVD algorithm [8] | $0.8\tilde{N}\tilde{S}$ flops + $O(\tilde{S}^3)$ |
| Max-Min-SNR alg. [11] | $1.6\tilde{N}\tilde{L}^2$ flops |
| CED algorithm | $\tilde{N}$ flops + $O(\tilde{N}\log(\tilde{N}))$ |
| Proposed GED alg. | $\tilde{N}$ flops + $O(\tilde{N}\log(\tilde{N}))$ |

Note that all of the proposed detection algorithms, (i.e., edge, reference white sub-band and generalized energy detections) do not depend on the phase of the received signal, and are not sensitive to carrier frequency offset and small to medium ACI signals. Therefore, the effects of carrier frequency offset, phase offset and adjacent channel interference on the performance of the proposed detection algorithms can be studied like that of [11]. The detailed analysis is omitted for conciseness.

## VIII. CONCLUSIONS

This paper considers wide-band spectrum sensing and throughput optimization for CR networks under noise variance uncertainty. It is assumed that the considered wide-band contains one or more white sub-bands. Under this assumption, we consider throughput maximization of the CR network while appropriately protecting the primary network. We address this problem as follows. First, we propose novel ratio based edge detector. Second, we employ a simple energy comparison approach to choose one reference white sub-band. Third, we propose novel GED for examining each of the target sub-bands by exploiting the noise information of the reference white sub-band. Finally, we optimize the sensing time for maximizing the throughput of the CR network by utilizing the $P_d$ and $P_f$ expressions of the GED. The proposed GED does not suffer from SNR wall and outperforms existing signal detectors. The relationship between the CED and the proposed GED is quantified analytically. The sensing time optimization problem is shown to be a concave maximization problem where its solution is obtained by simple bisection search approach. The analytical $P_d$ and $P_f$ expressions of the proposed algorithms are confirmed by computer simulation. The proposed algorithms are robust against noise variance uncertainty, carrier frequency offset and moderate ACI signals.

## APPENDIX A
### DERIVATION OF (12)

After some straightforward steps, the entries of $\boldsymbol{\Psi}$ becomes

$$\boldsymbol{\Psi}_{(1,1)} = N_{eh}(\mathrm{E}\{\hat{M}_v \hat{M}_v^H\} - M_v^2) = M_{v4} - M_v^2$$
$$\boldsymbol{\Psi}_{(2,2)} = N_{eh}(\mathrm{E}\{\hat{M}_{\tilde{v}} \hat{M}_{\tilde{v}}^H\} - M_{\tilde{v}}^2) = M_{\tilde{v}4} - M_{\tilde{v}}^2$$
$$\boldsymbol{\Psi}_{(1,2)} = \boldsymbol{\Psi}_{(2,1)} = 0 \qquad (44)$$

where $M_{v4}$ and $M_{\tilde{v}4}$ are the fourth moments of $v[n]$ and $\tilde{v}[n]$ respectively, and the last equality is due to the fact that $v[n]$ and $\tilde{v}[n]$ are uncorrelated. By applying the moment definition in [28], the following moments are obtained

Under $\mathrm{H}_{0e}$ $\begin{cases} M_{\tilde{v}} = M_v = \sigma^2 \\ M_{\tilde{v}4} = M_{v4} = 2\sigma^4 \end{cases}$

Under $\mathrm{H}_{1e1}$ $\begin{cases} M_{\tilde{v}} = \sigma^2, & M_v = (1+\gamma_e)\sigma^2 \\ M_{\tilde{v}4} = 2\sigma^4, & M_{v4} = 2(1+\gamma_e)^2\sigma^4 \end{cases}$

Under $\mathrm{H}_{1e2}$ $\begin{cases} M_v = \sigma^2, & M_{\tilde{v}} = (1+\gamma_e)\sigma^2 \\ M_{v4} = 2\sigma^4, & M_{\tilde{v}4} = 2(1+\gamma_e)^2\sigma^4. \end{cases}$

Substituting these expressions and (44) into (10) gives (12).

## APPENDIX B
### OPTIMAL SOLUTION OF (43)

The first and second derivatives of $\tilde{f}(T_o)$ (43) with respect to $T_o$ are given by

$$\tilde{f}'(T_o) = \sum_{k=1, k \neq i}^{S_{sb}} \psi_k \left[ \frac{a_k}{\sqrt{2\pi T_o}} \exp\left(-\frac{(a_k\sqrt{T_o}+b_k)^2}{2}\right) \times \right.$$
$$\left. \left(1 - \frac{T_o}{T_f}\right) - \frac{1}{T} \mathrm{erf}\left(\frac{a_k\sqrt{T_o}+b_k}{\sqrt{2}}\right) \right] - \frac{\tilde{\psi}_k}{T_f}$$

$$\tilde{f}''(T_o) = \frac{d\tilde{f}'(T_o)}{dT_o}$$
$$= \sum_{k=1, k \neq i}^{S_{sbg}} -\psi_k \frac{a_k}{\sqrt{2\pi T_o}} \exp\left(-\frac{(a_k\sqrt{T_o}+b_k)^2}{2}\right) \left[\frac{2}{T_f}\right.$$
$$\left. + \left(\frac{a_k}{2\sqrt{T_o}}(a_k\sqrt{T_o}+b_k) + \frac{1}{2T_o}\right)\left(1 - \frac{T_o}{T_f}\right)\right].$$

As can be seen from these expressions, for the practically relevant regions $0 \leq T_o \leq T_f$ and $a_k\sqrt{T_o}+b_k \geq 0$ (i.e., $P_{fk} < 0.5$), $\tilde{f}'(T_o)$ is a decreasing function and $\tilde{f}''(T_o) \leq 0$. Thus, in these regions, $\tilde{f}(T_o)$ is a concave function. Furthermore, since $\tilde{f}'(T_o) = 0$ exists, the optimal solution of (39) is given by $T_o^\star = \tilde{f}'(T_o) = 0$ which can be computed numerically by simple bisection search method [23], [24].